# DISCOVERY OF A z=2.76 DUSTY RADIO GALAXY?[1]

ARJUN DEY & HYRON SPINRAD[2 3]
Astronomy Dept., 601 Campbell Hall, U. C. Berkeley, CA 94720

AND

MARK DICKINSON[2 3]
Space Telescope Science Institute, Baltimore, MD 21218

## ABSTRACT

We report the discovery of a z=2.76 radio galaxy, MG1019+0535, with an unusual spectrum. Ly$\alpha$, which usually dominates the spectra of high redshift radio galaxies, is very weak and the strongest lines in the spectrum are CIV$\lambda$1549 and HeII$\lambda$1640. We speculate that dust is responsible for attenuating the Ly$\alpha$ line. This object provides more evidence for dust formation at early epochs.

The optical counterpart of the compact radio source has a double morphology which is orthogonal to the radio source axis and most likely due to the superposition of two distinct objects which may or may not be physically related. There is only one detected line from the brighter component (B) which is close in observed wavelength to the redshifted HeII$\lambda$1640 line from the z=2.76 galaxy (A). We discuss whether B is at the same redshift as A or whether it is a foreground galaxy at z=0.66. We also report the serendipitous discovery of a very red ($R-K>7$) field object.

## 1. INTRODUCTION

The identification of the optical counterparts of faint radio sources remains one of the most effective methods of finding galaxies at high redshifts (*e.g.* Minkowski 1960, Lacy *et al.* 1994), primarily because all powerful radio galaxies have strong emission line spectra. The strongest line typically seen in the rest frame UV spectra of z>1.8 radio galaxies is Lyman $\alpha$, a fact which led to early interpretations of high redshift radio galaxies as protogalaxies. Although we now believe that the luminous Ly$\alpha$ nebulae in high redshift radio galaxies are powered by their AGN rather than by stars, the discovery of strong Ly$\alpha$ emitters at high redshift is still regarded as a possible protogalactic signature (Partridge & Peebles 1967) and has prompted deep Ly$\alpha$ imaging searches for primeval galaxies. Thus far, these searches have all been unsuccessful (see Djorgovski *et al.* 1993 for a recent review). One possible reason for this is that dust suppresses the Ly$\alpha$ emission from protogalaxies; indeed, even small dust optical depths can completely extinguish the Ly$\alpha$ emission (Hummer & Kunasz 1980, Neufeld 1990).

Is there any evidence for dust at high redshift? Although there is weak statistical evidence for dust in damped Ly$\alpha$ systems (*e.g.* Fall & Pei 1993), more direct evidence has been found in the z=2.28 ultraluminous *IRAS* galaxy F10214+4724 (Rowan-Robinson *et al.* 1991) and, more recently, in the z=2.34 steep spectrum radio galaxy TX0211−122 (van Ojik *et al.* 1994). Both these objects show very weak Ly$\alpha$ lines and in addition F10214 is a powerful infrared emitter. In this paper we report the discovery of a z=2.76 radio galaxy with weak Ly$\alpha$. It has unusual spectroscopic and morphological properties and may be the most distant dusty galaxy discovered to date.

Over the last few years we have been pursuing the optical identification of a sample of 216 radio sources selected in a systematic manner from the 5GHz MIT/Greenbank survey (MG I) of Bennett *et al.* (1986). (Several other optical studies of radio flux density selected samples are presently in progress; see McCarthy 1993 for a recent review.) Our sample is selected based on radio source angular size ($\theta \leq 10''$) and radio spectral index ($\alpha^{5\mathrm{GHz}}_{1.4\mathrm{GHz}} \geq 0.75, F_\nu \sim \nu^{-\alpha}$) which biases the sample toward finding galaxies at high redshift. In addition, we only select sources which have simple (*i.e.* unresolved, double or triple) radio morphologies. The details of and results from our survey will be discussed in a future paper; some preliminary results are described in Spinrad *et al.* (1993). The galaxy discussed here has the second highest redshift in our sample.

Throughout this paper we use $H_\circ = 50$ km s$^{-1}$Mpc$^{-1}$ and $q_\circ = 0$ which implies an angular scale of 13.5 kpc/$''$ at z=2.76. For $H_\circ = 50$ km s$^{-1}$Mpc$^{-1}$ and $q_\circ = 0.5$, the corresponding scale would be 7.5 kpc/$''$.

## 2. OBSERVATIONS

MG1019+0535 (J2000; 1016+0549 B1950) was observed on numerous occasions over the past 2 years at the KPNO Mayall 4m and the Lick Observatory Shane 3m telescopes. The faint ($R \sim 23$) optical counterpart to this radio source was first identified on images taken with the 4m in April 1990; deeper images were obtained in better seeing with the 4m in April 1992. Spectra taken with the 4m and RC spectrograph in November 1990 and with the lens/grism spectrograph on the Lick 3m in December 1990 and January 1991 showed the CIV$\lambda$1549 and HeII$\lambda$1640 emission lines at a redshift $z = 2.76$, but Lyman $\alpha$ $\lambda$1216, normally by far the strongest UV spectral feature in high redshift radio galaxies, was only weakly detected. We considered this to be sufficiently unusual as to doubt the redshift until further observations obtained at Lick with the Kast spectrograph in February 1994 confirmed the redshift with a secure detection of CIII]$\lambda$1909. These previous optical observations have largely been superceded by the Keck data presented below; we therefore discuss only these latter data here.

[1]Based on observations at the W.M. Keck Observatory
[2] Visiting Astronomer at Kitt Peak National Observatory, National Optical Astronomy Observatories, which is operated by Association of Universities for Research in Astronomy, Inc., under contract by the National Science Foundation.
[3]Visiting Astronmer at the Very Large Array of the National Radio Astronomy Observatory, which is operated by Associated Universities, Inc., under contract by the National Science Foundation.

We observed MG1019+0535 at the Cassegrain focus of the 10m W. M. Keck Telescope using the Low Resolution Imaging Spectrometer (LRIS) on U.T. 1994 March 14 and 15. The detector is a Tektronix $2048^2$ CCD with 24$\mu$m pixels corresponding to a scale of 0$''$.214 pix$^{-1}$. Our observing run was one of the first Keck LRIS science runs and there were several problems with the instrument and detector: the readout noise of both amplifiers was excessively high ($25e^-$ on the blue side and $65e^-$ on the red side) and the red side showed non-linearity at moderate count levels which resulted in difficulties in flux calibration at wavelengths longer than 7300Å. The typical seeing during our observing run was 0$''$.8 – 1$''$.0.

Three direct images of MG1019+0535 (total $t_{exp}$ = 1300$s$) were obtained through a Cousins $R$ (3mm OG570 + 3mm KG3; $\lambda_{eff} \approx 6500$Å, $\Delta\lambda \approx 1400$Å) filter. The images were corrected for overscan bias, flat fielded using a median sky flat and coadded. The effective seeing in the coadded frame is 1$''$.0 and the $3\sigma$ detection limit in a 3$''$ diameter aperture is $R \approx$27.2 mag. The imaging data were calibrated using observations of the Landolt (1992) standard star field PG1525−071 and the spectrophotometric standard G193-74 (Oke 1990).

Five spectra (total $t_{\rm exp}$ = 7500$s$) of MG1019+0535 in the wavelength range 3900−8800Å were obtained through a 1$''$ slit in PA=36° using a 300 l/mm grating ($\lambda_b$ = 5000Å, resolution FWHM≈10Å). The slit was aligned on the object by using a blind offset from a bright offset star which lies 26$''$.6 west and 28$''$.0 north of the MG source. The slit PA was chosen to trace the axis through the two components of the galaxy; it was also parallel to the average parallactic angle during our observation. Hence, although the spectroscopic data were obtained under non-photometric conditions, we believe our relative spectrophotometry to be accurate for $\lambda <$ 7300Å. The data were corrected for overscan bias and flat fielded using internal lamps taken immediately following the observations. Spectra of the two components were extracted using parallel linear models for the centroid variation along the dispersion axis and were flux calibrated using observations of the spectrophotometric standards G193-74 (Oke 1990) and HZ44 (Massey *et al.* 1988). The standard stars were observed both with and without an OG570 filter in order to correct for the second order light contaminating the wavelength region $\lambda >$ 7500Å.

The field was imaged in the $K_S$ band (1.99–2.32$\mu$m) using the Infrared Imager (IRIM) on the KPNO 4m on UT 1994 February 25. IRIM employs a 256×256 pixel HgCdTe NICMOS-3 array with a scale of 0$''$.603/pixel at the 4m F/15 Cassegrain focus. Thirty frames were obtained, each an average of three 20 second exposures, with the telescope offset between frames in a non–redundant pattern. Array response variations were corrected using dome flat exposures. Sky frames were created for and subtracted from each image using a object–masked combination of eight temporally adjacent exposures. The resulting frames were spatially oversampled by a factor of four, co–registered to the nearest sub–pixel, and summed. The total exposure time in the resulting image is 1800 seconds, and the image PSF has a FWHM=1$''$.1. Sky conditions were judged to be photometric based on minimal frame–to–frame zeropoint variations and the excellent quality of the sky subtraction; the zeropoint uncertainty is small compared to the random error in the photometry on this faint source. Observations of faint UKIRT infrared standards (Casali & Hawarden 1992) provided photometric calibration, under the assumption that $K$=$K_S$ for these stars (reasonable given their nearly uniformly zero $H-K$ colors). The reduction of all our imaging and spectroscopic data were done using the IRAF package.

MG1019+0535 was observed in snapshot mode using the VLA in the A-array configuration on 12 August 1991 at 8.44GHz and 1.4GHz. The exposure time in each band was roughly 7 min. We observed the VLA calibrator 0922+005 for phase calibration and 3C286 for flux calibration. Calibration and analysis of the radio data were carried out using the AIPS package. The data were self calibrated in three iterations, the first two iterations using only phase self calibration and the last using amplitude self calibration.

## 3. Results

Figure 1 shows a detail from our Keck $R$-band image of MG1019+0535. The galaxy has two components separated by 1$''$.5 in PA=36° of which the southern, fainter component ('A') is probably associated with the radio source. In §4.2, we discuss the interpretation of this double morphology in some detail. Here, we simply note that the $R$-band image is well fit by the superposition of two distinct model galaxies with elliptical isophotes. A measurement of the $R$ magnitude of each component was obtained after subtracting out the best fit model to the other component. We present the photometry in Table 1, including a pure–continuum $R$ magnitude for object A after removing a 17% emission line contribution derived from our spectroscopy.

A panel showing both the $R$ and $K_S$ images is shown in Figure 2 (Plate). The $K$ image of MG1019+0535 is crudely consistent with the two component morphology of the $R$ image. Although the signal–to–noise ratio in the $K$ frame is too poor to allow accurate division of the total flux between the two components A and B, we have attempted a rough estimate by using the two elliptical model constructed for the $R$ image. We scaled the fits to the two $R$-band components independently and subtracted them from the $K$ image until the residuals were minimized. The resulting magnitudes for the individual components are listed in Table I with an estimate of the errors. The $K_S$ band is free of strong emission lines at the redshift of the radio galaxy.

The VLA map of the radio source is shown in Figure 3. The radio source has a triple structure with two bright lobes ($f^{\rm east}_{8.44{\rm GHz}}$ = 24mJy, $f^{\rm west}_{8.44{\rm GHz}}$ = 27.5mJy) separated by about 1$''$.3 and a faint core. The spectral index integrated over the radio source is $\alpha^{8.44{\rm GHz}}_{1.4{\rm GHz}} = -1.0$. The radio axis PA of 103° is roughly orthogonal to the line joining the two optical components. This is unusual for high redshift radio galaxies which generally show alignments between their rest frame UV morphologies and their radio axes (McCarthy *et al.* 1987, Chambers *et al.* 1987). Table 1 lists the radio and optical positions and flux densities of the various components of MG1019+0535.

The cross in Figure 1 marks our best estimate of the optical position of the radio core based on 5 astrometric stars in the field. Using different combinations of astrometric stars resulted in slightly different positions for the core position (with slightly larger errors) but these all appear to scatter close to the position of the optical component **A**. Therefore we conclude that the radio source is most likely to be associated with component **A**. If the core coincides with the optical component **A**, then the extended radio structure is entirely confined within the optical extent of the galaxy. Component **A** may be slightly extended in the east-west direction implying that it may be aligned with the radio source axis.

The LRIS spectra of the two optical components are shown in Figure 4. The central wavelengths, fluxes, equivalent widths and full widths at half maximum of each emission line in the observed frame are listed in Table 2. The fluxes listed are those measured in a 1″×1.″7 aperture. Component **A** has strong and narrow associated line emission from Ly$\alpha$, NV, CIV, HeII and CIII] at a (weighted) mean redshift z=2.765±0.002 (Fig. 4*a*). The Ly$\alpha$ line is redshifted with respect to the CIV, HeII and CIII] lines by $\sim$500 km s$^{-1}$; this is similar to offsets observed in other high redshift radio galaxies and quasars. In addition, there is a tentative detection of OVI$\lambda$1034 at $\lambda_{obs} = 3903$Å. This feature is at the very edge of our spectral window where the instrumental sensitivity is decreasing and therefore should be regarded as uncertain.

The CIV, HeII and CIII] lines are broader than the Ly$\alpha$ and NV lines. For reference, the night sky lines of [OI]$\lambda$5577 and [OI]$\lambda$6300 have a FWHM $\approx$10Å in our spectrum. This implies that the Ly$\alpha$ and NV lines are unresolved, whereas the HeII line is broadened by a velocity dispersion of $\sim$400 km s$^{-1}$.

Component **B** has a single, unresolved, moderately strong line at 6192Å, separated by only 17Å in observed wavelength from the redshifted HeII line in component **A**. Its continuum spectrum appears to have a discontinuity at $\lambda \approx 4500$Å below which there is little detectable flux. The spectrum appears to rise longward of $\lambda \approx 6300$Å. The drop in the flux at $\lambda \gtrsim 7600$Å is not real and probably due to instrumental effects (*e.g.* problems with the red side CCD readout amplifier, the decreasing CCD sensitivity at long wavelengths and the increased read noise on the red side). The optical spectrum of **B** appears redder than that of **A**, although its $R-K$ color is slightly bluer.

Since only one strong line is detected from component **B**, we cannot be certain of its redshift. Three possibilities are (1) that the line is Ly$\alpha$, (2) the line is [OII]$\lambda$3727 which is one of the strongest lines in low redshift star forming galaxies, and (3) it is HeII$\lambda$1640. The last possibility is based primarily on the proximity of the line in **B** to the HeII$\lambda$1640 line in **A**. In addition, whether or not **B** is physically associated with component **A**, the unusual spectrum of **A** still requires an explanation. These issues are discussed in §4.2.

## 4. Discussion

### 4.1 The Unusual Spectrum of Component A and the Evidence for Dust

The most unusual aspect of the spectrum of component **A** is that the Ly$\alpha$ line is very weak compared to that observed in most high redshift radio galaxies. Table 3 compares the various line strengths (relative to HeII$\lambda$1640) with the corresponding ratios in a few 'canonical' objects. The composite radio galaxy spectrum was constructed from 3CR and 1 Jy–class galaxies with 0.1<z<3 by McCarthy (1993). The UV spectrum of NGC1068 was obtained by Kriss *et al.* (1992) using the Hopkins Ultraviolet Telescope. The only other reported examples of high redshift objects with suppressed Ly$\alpha$ are the dusty, ultraluminous *IRAS* galaxy F10214+4724 at $z$=2.286 (Rowan-Robinson *et al.* 1991, Elston *et al.* 1994), and a $z$=2.34 radio galaxy TX0211−122 (van Ojik *et al.* 1994). TX0211−122 has been argued to be a dusty galaxy based on the similarity of its rest frame UV spectrum to that of F10214+4724. For comparison, Table 3 also lists the line ratios predicted by the photoionization code CLOUDY (Ferland 1994) for two values of the ionization parameter ($U = Q(\mathrm{H}^\circ)/4\pi r^2 N_e$) of −1.8 and −2.8. In both cases a power law input spectrum of $S_\nu \sim \nu^{-1.5}$ and a density of 10 cm$^{-3}$ was used.

Although the CIV/HeII/CIII] ratios observed in object **A** are similar to those in McCarthy's composite radio galaxy spectrum, the NV/HeII and Ly$\alpha$/HeII ratios in **A** are respectively factors of 2 and 10 times lower than those in the composite spectrum. The Ly$\alpha$/HeII ratio varies only slowly with ionization parameter (*e.g.* Ferland & Netzer 1983) and therefore it is unlikely that the observed ratio is due to differences in $U$, although varying the shape of the input spectrum may change the observed ratio. It is interesting to note that the fact that the CIV/HeII/CIII] ratios are all roughly unity implies that the ionization parameter is about 0.01 (*e.g.* Ferland & Netzer 1983), which is similar to that derived for most radio galaxies. We do not see the anomalously high NV strength observed in the spectra of F10214+4724 and TX 0211−122 (van Ojik *et al.* 1994).

In addition to component **A**'s unusual line ratios, its Ly$\alpha$ luminosity is also intrinsically weak ($L_{\mathrm{Ly}\alpha} = 1.6 \times 10^{43}$ erg s$^{-1}$). McCarthy (1993) presents an empirical correlation between emission line luminosity and 1.4 GHz radio power. Using canonical values for the Ly$\alpha$/[OII] ratio of 5 to 8 (also from McCarthy 1993) to convert his relationship for our purposes, and assuming negligible contribution from HeII$\lambda$1215 to the observed line flux from MG 1019+0535 (see below), we find that its Ly$\alpha$ luminosity is a factor of 30 to 50 below the empirical correlation. We note, however, that the scatter in the $L_{[\mathrm{OII}]}$ vs. $P_{1.4\mathrm{GHz}}$ relation is very large, being almost an order of magnitude. The Ly$\alpha$ line in MG 1019+0535 does not appear to be spatially extended in our two–dimensional spectra, either in our Lick/KPNO data with the slit oriented along the radio axis (approximately PA=90°) or in our Keck spectra with the slit oriented along the line joining the two components (PA=36°). This is in sharp contrast to most powerful, high redshift radio galaxies, where the Ly$\alpha$ emission is both strong and spatially extended (generally along the radio axis) over many tens of kpc (*e.g.* 3C294 McCarthy *et al.* 1990).

The most straightforward method of destroying Ly$\alpha$ is using dust. Since Ly$\alpha$ is a resonance line, multiple scatterings by ground-state hydrogen atoms significantly increase the path length traversed by a Ly$\alpha$ photon in a nebula thus increasing the probability of absorption of the photon by dust (Hummer & Kunasz 1980, Neufeld 1990). In MG1019+0535, a modest $E(B–V)$ of 0.3 or 0.4 and a Galactic extinction curve will correct the observed NV$\lambda$1240/HeII$\lambda$1640 ratio to 0.44 – 0.51, a value similar to that measured for the 'average' radio galaxy (Table 3). This $E(B–V)$ estimate assumes that the dust extinction is mostly external to the line emitting region (*e.g.* a foreground dusty screen), but if the relative velocity between this dusty gas cloud and the line emitting gas cloud is negligible and the velocity gradients within the two clouds are comparable, then most of the emitted Ly$\alpha$ photons will be trapped in the dust cloud. A similar quantity of dust intermixed with the line emitting gas could also be responsible for attenuating the Ly$\alpha$ line as long as the velocity gradients in the line emitting gas are small so as to prevent most of the Ly$\alpha$ photons from escaping. This amount of extinction is more than sufficient to completely suppress the Ly$\alpha$ emission relative to HeII$\lambda$1640. For example, in the recent calculation by Chen & Neufeld (1994), it is found that for even small total absorption optical depths, the fraction of photons that escapes in the center of the Ly$\alpha$ line can be more than an order of magnitude lower than the fraction that escapes at, for example, the position of the NV line 24Å away from the Ly$\alpha$ line center. In fact, it is extraordinary that we observe Ly$\alpha$ *at all* in distant galaxies. The geometry of the line emitting region and the spatial distribution of dust and dust to gas ratio in most objects must conspire to permit the ionizing radiation to reach a relatively dustless environment. Hence, although it is possible that there are other more involved methods of suppressing the Ly$\alpha$ line, the simplest method which also is consistent with the observed spectrum is extinction by dust. We therefore conclude that MG1019+0535 is probably a dusty galaxy.

Although the determination of the extinction using the NV/HeII and NV/CIV ratios is uncertain due to the intrinsic variation of these ratios that is observed in radio galaxies, the estimate of $E(B-V)$=0.3 is comparable to other reddening estimates for high redshift radio galaxies. For example, McCarthy *et al.* (1992) obtained an estimate of $E(B\text{-}V)\approx 0.1$ for two z=2.4 radio galaxies by using the Ly$\alpha$/H$\alpha$ line ratio. Both of these galaxies have very large Ly$\alpha$/CIV ratios (>5) and it is therefore likely that the reddening is larger in MG1019+0535 which has Ly$\alpha$/CIV $\approx$ 1.

We can estimate an upper limit to the continuum reddening by requiring the intrinsic continuum to be no bluer than flat spectrum (i.e. $f_\nu \sim \nu^\kappa, \kappa \leq 0$), the bluest color observed for distant radio galaxies thus far (e.g. Eisenhardt and Dickinson 1992, Eales *et al.* 1993). With this assumption, the maximum allowable extinction would be $E(B-V)\leq 0.43$. If the continuum and lines have a common extinction, the reddening is constrained to be $0.3 \lesssim E(B-V) \lesssim 0.43$. However, the narrow line emission and continuum components may originate in physically distinct regions as is probably the case for F10214+4724 where the extinction derived from the emission lines would deredden the continuum to be bluer than any observed galaxy or AGN spectrum (Elston *et al.* 1994).

If we assume that the extinction in MG1019+0535 is $E(B-V)$=0.3, then for the mean Galactic extinction curve of Savage & Mathis (1979) the extinction coefficients for the $R$ and $K$ filters in the rest frame of MG1019+0535 are $A_{R,z=2.765} = 8.08E(B-V)$ and $A_{K,z=2.765} = 2.92E(B-V)$. The dereddened continuum colour (after correcting for line contamination to the $R$ photometry) would then be $(R-K)$=2.36.

An observation of the HeII$\lambda$4686 line (which is redshifted into the $H$ band at 1.5454$\mu$m) would provide an independent measure of extinction. The Case B recombination values for this ratio lie between 5.4 and 9.0 for 3000K<T<50000K and 100cm$^{-3}$< $n_e$ < 10$^7$cm$^{-3}$ (Hummer & Storey 1987). Hence the predicted flux of the HeII$\lambda$4686 line for the extreme ($T$=50000K, $n_e$=10$^7$cm$^{-3}$) case is $\approx 10^{-17}$ erg cm$^{-2}$ s$^{-1}$; this is a lower limit if dust extinction is important.

Other evidence for dust in high redshift galaxies has been presented by McCarthy *et al.* (1992) and Eales & Rawlings (1993) who measured the Ly$\alpha$/H$\alpha$ and Ly$\alpha$/H$\beta$ ratios in a few high redshift radio galaxies. These line ratios are measured to be much lower than predicted by photoionization models, implying the existence of dust. The one high redshift object which shows convincing evidence for dust is the IRAS source F10214+4724; this object has suppressed Ly$\alpha$, a large far infrared luminosity presumably due to dust emission, and large amounts of molecular gas (Rowan-Robinson *et al.* 1991, Solomon *et al.* 1992). If MG1019+0535 were to have the same infrared luminosity as F10214+4724, its larger redshift would cause the measured infrared flux densities to be ~1.5 – 2 times fainter, rendering it undetectable in the *IRAS* database.

Table 3 shows that the HeII$\lambda$1640 emission line is comparable in strength to Ly$\alpha$ (Fig. 4*a*, Table 2). The HeII$\lambda$1640 line is the "Balmer" $\alpha$ (3→2) transition of the HeII ion. The corresponding HeII Balmer $\beta$ transition is at 1215.17Å and therefore contributes to the measured hydrogen Ly$\alpha$ flux. In most known high redshift galaxies the HeII$\lambda$1640 line is very weak relative to Ly$\alpha$ and this contribution is neglible. However, in the present case the observed HeII$\lambda$1640 line is very strong and the contribution needs to be considered. For Case B recombination, the HeII$\lambda$1215/HeII$\lambda$1640 ratio ranges between 0.27 and 0.33 for temperatures 5000K< $T$ <20000K and densities 100cm$^{-3}$ < N$_e$ < 10$^4$cm$^{-3}$ (Hummer & Storey 1987, Osterbrock 1989). As a result, the "true" Ly$\alpha$/HeII$\lambda$1640 ratio in MG1019+0535 probably lies between 0.67 and 0.72.

Note that the strong HeII$\lambda$1640 line implies that the nebular gas has a high degree of ionization which lends some credence to our tentative identification of the OVI line. If real, the OVI/HeII ratio of ~ 0.4 is similar to the ratio of 0.6 measured in the composite radio galaxy spectrum of McCarthy (1993). In comparison, the HUT spectra of NGC4151 and NGC1068 show OVI *stronger* than HeII$\lambda$1640 (Kriss *et al.* 1992*a*,*b*).

### 4.2 Morphology and Photometry

The most striking aspect of Figure 1 is the double morphology of the system which is reminiscent of low redshift double galaxy systems (*e.g.* 3C89, 3C169.1, 3C433). However, the fact that we detect only one strong line in the spectrum of component B raises the question as to whether (*i*) A and B are at the same redshift (z=2.76) and separated spatially by ≈20 kpc (in projected distance) and in velocity by only ≈800 km s$^{-1}$ or (*ii*) whether the two components are unrelated and at two different redshifts. In this latter case, the strong emission line seen in the spectrum of component B is most likely to be [OII]$\lambda$3727 implying that B is at z=0.66.

The $K$ magnitude for component A alone is faint for its redshift, about 1 magnitude less luminous than the median value implied by the $K$ vs. z diagrams in McCarthy 1993 and Eales et al. 1993. Given the observed scatter in radio galaxy $K$ magnitudes at these large redshifts, however, this deviation from the typical values is perhaps not unusual. If the radio galaxy were instead considered to consist of components A plus B together, its $K$-luminosity would be typical for its redshift. The color ($R-K \approx 3.9$ for component A alone, $R-K \approx 3.1$ for the entire system A+B) is not unusual for radio galaxies at z>2. If we apply the extinction correction implied by $E(B-V) = 0.3$ to the continuum of object A, its $K$ luminosity would be consistent with the $K$ vs. $z$ prediction, although the resulting corrected $R$ magnitude would then be ~1 to 2 magnitudes brighter than the norm.

In what follows, we discuss the evidence for the 'discrepant redshift' and 'same redshift' cases.

#### *4.2.1 A Chance Superposition....*

The main argument for different redshifts for A and B is that while component A shows emission lines from Ly$\alpha$, CIV, HeII and CIII], component B has only one line. In our experience, the detection of a single strong line over a large searched range in wavelength usually implies that the line is either Ly$\alpha$ or [OII]$\lambda$3727. It is unlikely that the single line from B is Ly$\alpha$ because (*i*) there is no drop in the continuum shortward of the line which would be expected due to Ly$\alpha$ forest absorbers and (*ii*) there is some detectable continuum shortward of 4650Å which would correspond to 912Å in the rest frame. On the other hand, it is entirely plausible that the line is [OII]$\lambda$3727 at z=0.66. In this case, the luminosity of the line is large ($L_{[\rm OII]} \approx 8 \times 10^{40}$ ergs s$^{-1}$), but not unprecedented for star forming galaxies. The rest frame equivalent width of the line is $W_{[\rm OII]} \approx 26$Å which is well within the range of equivalent widths measured in the integrated spectra of nearby galaxies (*e.g.* Kennicutt 1992*a*, Colless *et al.* 1990, 1994). In addition, the single emission line from B is unresolved, which implies that the width of the line is probably less than 240 kms$^{-1}$. Hence it is unlikely that B is a luminous AGN.

If B is at $z = 0.66$, then the rising continuum redward of $\lambda_{rest} \approx$3800Å ($\lambda_{obs} \approx$6300Å, see §3 and Fig. 4*b*) is typical of star forming galaxies, in which it is dominated by the continuum from intermediate age stars (in contrast to the "true" 4000Å break feature seen in old stellar populations). The break amplitude for B is $D_\nu(4000)$=1.56, quite typical for star forming galaxies at z=0.6 (Bruzual & Charlot 1993). The $R-K$ color of object B is also consistent with that of an actively star forming galaxy (e.g. a late–type spiral) at $z = 0.66$.

Ignoring the absence of other strong emission lines for the present (see §4.2.2), it is of note that almost all of the galaxies in the Kennicutt (1992*a,b*) galaxian spectral atlas with $W_{[\rm OII]} \geq 25$ are very late–type galaxies: except for 1 S0 which is a Seyfert, they all either have Hubble types of Sc or later, or are peculiar or interacting galaxies. If B is at $z$=0.66, then the $R$-band filter samples the continuum light between ≈3300–4800Å. The resolved morphologies of most late–type galaxies are more diffuse and much less nucleated at these wavelengths than the morphologies of early–type galaxies (*e.g.* Schweizer 1976). Although our ability to distinguish small scale morphological details at this redshift is degraded by our resolution (the diameter of our seeing disk corresponds to ≈7.5 kpc at $z$=0.66), simulations show that the overall surface brightness profile of component B is better fit by an exponential disk convolved with the seeing than a deVaucouleurs profile. Therefore the morphology of component B is consistent with that of a $z$=0.66 late–type spiral.

There are two arguments against the low redshift hypothesis for B. Firstly, we do not detect other strong lines (such as [OIII] and H$\beta$) that are commonly observed in the spectra of low redshift emission line galaxies. This is discussed in further detail in §4.2.2. Secondly, low redshift emission line galaxies are not known to have a spectral discontinuity at $\lambda_{rest} \approx$2700Å ($\lambda_{obs} \approx$4500Å; see Fig. 4*b*). The UV spectra of early–type galaxies can be steeply declining in this region. However, the optical spectrum, the optical morphology and the optical–IR color suggest instead that B would be a late–type spiral which would therefore have a much flatter UV spectral energy distribution.

If the double morphology of the system is due to a chance superposition, is this an extremely unlikely event? Let us consider the probability that we would find a chance superposition between a radio source and an unrelated field galaxy along the same line of sight. There are 216 radio sources in our sample, of which we have imaged 100 to date. The surface density of galaxies with $R$ magnitudes between 18.0 and 23.5 is approximately $\Sigma = 3\times10^{-3}$ arcsec$^{-2}$ (Metcalfe *et al.* 1991; galaxies brighter than $R$=18 contribute insignificantly to the counts). Therefore, the *a priori* expected number of superpositions between unrelated field galaxies and radio sources within $\theta''$ of one of the radio sources in our sample is crudely $100\Sigma\pi\theta^2$, or ∼2 for $\theta$=1.″5. Hence it is not improbable that we should find such a superposition. Here we have neglected the angular correlation function of galaxies under the assumption that chance projections should be uncorrelated; including it into the calculation roughly doubles the probability. Gravitational lensing by the foreground galaxy could enhance the likelihood for such a projection, but we see no obvious evidence for this in the case of MG 1019+0535. In fact component A is rather underluminous at $K$ compared to other radio galaxies at similar redshift.

Using the redshift distribution of faint field galaxies from combined $I < 22.5$ samples of Lilly (1992) and Tresse *et al.* (1993) (which should be roughly equivalent to that for galaxies at $R \lesssim 23$), the probability that a given field galaxy would have a redshift placing [OII]$\lambda$3727 within 17Å of a strong line in the radio galaxy (Ly$\alpha$, CIV, HeII or CIII]) is 0.036. Of the galaxies in those surveys for which the spectroscopic observations cover the [OII] line, ∼78% have detected emission. Thus the combined expectation value for the observed superposition using the joint angular proximity, wavelength coincidence, and emission line fraction criteria is roughly $2 \times 0.036 \times 0.78 = 0.056$.

In summary, it is plausible that B is a foreground object, although it would have somewhat unusual spectral properties. The similarity of the emission line wavelengths would, in this scenario, be a somewhat remarkable coincidence, but not an infinitesimally likely one.

#### *4.2.2 ....Or A Double Galaxy System?*

An important argument against the $z = 0.66$ scenario for component B is the absence of [OIII]$\lambda$5007,4959 and H$\beta$ emission, which should appear in our spectral window. If B is at z=0.66, the rest frame equivalent width of the strong emission line in B is 26Å. Although this is fairly typical of widths observed in low redshift galaxies (*e.g.* Kennicutt 1992*a*, Colless *et al.* 1990,1994), galaxies that exhibit similar $W_{[\rm OII]}$ are usually star forming galaxies or AGN. In either case the [OIII] lines would also be strong and easily detectable. For star forming galaxies, the observed [OIII]/[OII] ratio exhibits a large range from 0.1 to 1 and in particular, for galaxies which have $W_{[\rm OII]} \geq 25$Å, the ratio is usually observed to be greater than 0.2 (*e.g.* Kennicutt 1992*a*). For most AGN other than LINERs, this ratio is typically greater than 1 (*e.g.* Baldwin, Phillips & Terlevich 1981). However, based on the spectrum of B, the limit on [OIII]/[OII]<0.03 implies that B is an unusual object if it is at the lower redshift.

The spectrum of a z=0.544 field object that we obtained serendipitously with the spectrum of MG1019 provides a useful comparison. This serendipitous object lies about 44″ east and 60.″5 north of MG1019+0535. Its spectrum shows strong [OII]$\lambda$3727 emission ($f_{obs}$ = $4.3 \times 10^{-17}$erg s$^{-1}$ cm$^{-2}$,$W_{\lambda,rest}$ = 65Å) but *also* exhibits strong [OIII]$\lambda$4959,5007 and H$\beta$ emission with the ratios [OIII]$\lambda$5007/[OII] = 1.33 and H$\beta$/[OII] = 0.47. Although the serendipitous galaxy was not centered on the slit and therefore the line ratios may only be representative, it is important to note that we were easily able to detect the [OIII] and H$\beta$ lines in this galaxy but not in component B.

An additional clue is obtained from comparing the flux ratio across the Ly$\alpha$ line in the two components. For component A, the Ly$\alpha$ continuum (forest) discontinuity ($f_\lambda(1300-1400)/f_\lambda(1100-1200)$) is roughly 1.5±0.1. However, B has very little flux below $\lambda_{obs} \sim 4500$ (Fig. 4*b*). If B is at z=2.76, this translates into a 'Ly$\alpha$ forest' continuum discontinuity of about 4.5±0.5. This is extremely large; typical values of the forest discontinuity at z=2.7 lie between 1.2 and 1.7 (Schneider, Schmidt & Gunn 1991). Hence, if this break is real, it can only be explained by invoking some strongly absorbing gas such as would result in a strong damped Ly$\alpha$ system or a BAL-type system. Unfortunately we can neither confirm nor rule out whether this break is an instrumental effect; a comparison with the serendipitous z=0.544 galaxy is inconclusive since its spectrum has very little detected continuum below 5000Å. However, as discussed in §4.2.1, a break of this amplitude is not usually seen in emission line galaxies.

If both A and B are at z=2.76, then the strongest UV line in the spectrum of B is HeII$\lambda$1640 which is unusual (perhaps unprecedented). However, there is no *a priori* physical reason why this situation could not occur. Simulations with CLOUDY suggest that the CIV/HeII ratio is fairly sensitive to the ionization parameter. Specifically, decreasing the ionization parameter from $\log(U) = -1.8$ to $\log(U) = -2.8$ for an $f_\nu \sim \nu^{-1.5}$ input spectrum on solar abundance gas changes the CIV/HeII ratio from 1.13 to 0.01 (Table 3). If we assume that the sole source of ionization in B is due to the AGN nucleus in A, then B would experience a highly diluted radiation field relative to that experienced by the ambient medium in A. Alternatively, the ionization parameter may be decreased in B relative to A by invoking a higher particle density for the ionized gas in B relative to A. The Ly$\alpha$ line from B is easily suppressed by dust. Although suppressing the CIV relative to HeII requires lowering the ionization parameter an order of magnitude from the 'canonical' model, $U$ must be decreased an additional order of magnitude ($\log(U)<-4$) to suppress the CIII] line relative to HeII.

Given that the Ly$\alpha$ line implies the presence of dust, it is suggestive that the double optical morphology of the system resembles that of nearby dusty radio galaxies such as Centaurus A (Baade & Minkowski 1954*b*, Burbidge & Burbidge 1959) and Cygnus A (*e.g.* Baade & Minkowski 1954*a*). Could it be that we are observing a dust lane in this high redshift galaxy? There are two arguments against this suggestion. Firstly, as noted above, we can model the observed structure well by using two separate objects with undistorted elliptical isophotes. The morphology is therefore consistent with the superposition of two separate, undistorted galaxies (without specifying their relative redshifts). This implies (invoking Occam's razor) that a dust lane is unnecessary in order to explain the morphology. Secondly, the optical astrometry places the radio core position very near component A. If the double optical system were indeed a single galaxy with a dust lane, one might naively expect the radio source to be centered between the two components and orthogonal to the radio axis, as is observed in Centaurus A (Wade *et al.* 1971, Meier *et al.* 1989).

If the two components are instead separate objects at the same redshift, then their projected separation is only 20 kpc and their observed difference in velocity is 800 kms$^{-1}$. This is of the order of cluster velocity dispersions. In fact, even larger relative velocities have been measured for double galaxies that are seen in association with radio sources. One such example is 3C169.1 (z=0.633), in which the two components of the double galaxy system are separated by only 3″ (27 kpc projected distance) and have a relative velocity of 1400 kms$^{-1}$! The observed morphology and kinematics of low redshift radio galaxies with double components have been argued by various authors to support the hypothesis that these systems are merging galaxies and that the AGN activity has been triggered by the merger (*e.g.* Heckman *et al.* 1986). With such a small projected separation and (if they are at similar redshifts) moderate relative velocity, it is possible that the the two optical components of MG1019+0535 are in the process of merging.

#### *4.2.3 An Extremely Red Object*

A visual comparison of the $R$ and $K$ images of the MG1019+0535 field reveals an extremely red object (D in Fig. 2), which is ≈ 10″ W of the radio source. While it is easily detected in the infrared image ($K$=18.26±0.12 in a 4″ diameter aperture), it is undetected in our deep Keck $R$ image. The object is nonstellar: it is spatially extended in $K$ with a deconvolved FWHM ≈ 0.″5. Although its proximity to the brighter galaxy C makes accurate photometry difficult in the $R$ image, a conservative estimate of its color is $R-K>7$. Without invoking dust extinction, such red colors are most naturally explained as resulting from the large K–corrections expected for relatively unevolved, early–type galaxies at $z>1$. The presence of a presumably foreground galaxy (C) close to the line of sight to D, however, could result in additional reddening. The bright $K$ magnitude of the object makes it unlikely that it is at the redshift of MG 1019+0535.

Similar very red objects have been discovered close to several powerful AGN at large redshifts. For example, McCarthy, Persson & West (1992) reported the discovery of very red objects ($R-K>6$, $K\sim 18$) near the radio galaxies 0406–244 (z=2.427) and 2025–218 (z=2.630). The field of B2 0902+34 (z=3.395, Lilly 1988) contains a $K=18.3$, $R-K=6.1$ object which lies 15″ from the radio galaxy (Eisenhardt & Dickinson 1992). Graham *et al.* (1993) discovered two red objects with identical magnitudes and colors ($K=18.4$, $R-K=6.2$) that lie 10″ and 20″ away from the high redshift galaxy 4C41.17 (z=3.80, Chambers, Miley & van Breugel 1990). The two $I-K\approx 6.5$ ($K=18.4$ & 18.7) objects reported by Hu & Ridgway (1994) are in the field of the z=3.79 quasar PC 1643+4631A; one of the objects is less than 20″ from the quasar.

Infrared surveys of blank fields have also discovered very red galaxies with bright $K$ apparent magnitudes (cf. Elston, Rieke & Rieke 1991). The field surface density of the reddest objects is $\lesssim 0.01/\Box'$ based on a recent reanalysis of the Hawaii Deep $K$-band Survey data by Hu & Ridgway (1993). In comparison, the detections summarized above imply that the surface density of very red objects in the fields around high redshift radio galaxies is unusually large: $\gtrsim 1/\Box'$. Obviously a more thorough statistical study of the presence of very red objects in the fields of high redshift AGN with a proper treatment of the field population is required before drawing any conclusions on the basis of the few fields mentioned here. If these objects are $z \gtrsim 1$ ellipticals, then an excess of these objects in the fields of high redshift AGN may imply that gravitational lensing is biasing our samples of the most distant radio galaxies. On the other hand, if they are at the redshift of the AGN and their emission is primarily starlight, these objects would provide stringent constraints on the epoch of galaxy formation. The bright $K$ magnitudes of the red objects, however, imply that this latter possibility is unlikely.

## 5. CONCLUSIONS

We have discovered a dusty z=2.76 radio galaxy as part of our ongoing study of radio sources from the MIT-Greenbank radio survey. The optical image shows two components separated by 1″.5. Although these may be physically related, there is circumstantial evidence that one of the components is a foreground object at z=0.66. Whether or not the two components are related, the spectrum of the high redshift object is extremely unusual: it has very weak Ly$\alpha$ and an observed HeII/Ly$\alpha$ ratio of $\approx$1. The most straightforward method of suppressing Ly$\alpha$ relative to HeII is by dust extinction of the resonance line; therefore MG1019 is a good candidate for a high redshift dusty radio galaxy. The strength of the HeII$\lambda$1640 emission line implies that HeII$\lambda$1215 might be an important contribution to the observed H Ly$\alpha$ line strength.

We have discussed the likelihood that the two optical components are at the same redshift or at two different redshifts. If they are at different redshifts, then the spectrum of B is unusual with a moderately luminous [OII]$\lambda$3727 line and a small [OIII]/[OII] ratio. However, the gross features of the spectrum (*i.e.* the strength of the single line and the shape of the continuum) appear to weigh in favor of the conclusion that component B is at z=0.66. If, however, they are both at z=2.76, then component B has an even more unusual spectrum with HeII$\lambda$1640 being the only strong UV line. We speculate that it is possible for such a situation to arise if the line emitting gas in B is ionized by the diluted radiation field of the AGN in A. In addition, the observed morphology may imply that we are observing an interacting system at z=2.76.

The discovery of dusty systems (or even galaxies with weak Ly$\alpha$) at high redshift has several important implications. Firstly, it implies that population synthesis modelling of the broad band colours of high redshift radio galaxies may not be straightforward since these models so far do not account for any reddening of the continuum by dust. Secondly, it is important to note that there are only three known high redshift (z>1.8) radio galaxies with weak Ly$\alpha$. This implies that either dusty high redshift radio galaxies are very rare objects and most high redshift radio galaxies are largely dust free, or the dusty objects are missing from our high redshift samples due to selection and observational biases. Thirdly, it provides further evidence that dust extinction may be the reason that primeval galaxy searches designed to look for strong Ly$\alpha$ line emission have been largely unsuccessful. Finally, the existence of dust at high redshifts raises the question of how and when this dust was formed. The presence of dust implies an early episode of star formation. This suggests that it is more appropriate to interpret the spectral energy distributions of high redshift radio galaxies in terms of an underlying old population and a superimposed burst population (cf. McCarthy 1993) than in terms of a single burst of star formation.

We are very grateful to B. Burke and S. Conner at MIT for introducing us to the MG sample and for supplying much of the astrometry for our candidate subset. We thank Joe Shields for interesting discussions and are extremely grateful to him for providing us with the predictions by CLOUDY. We would like to thank C. Steidel for assistance with the IR observations and W. van Breugel for help with the radio observations and reductions. We also thank Tom Bida, Wayne Wack and the W. M. Keck Observatory staff for invaluable assistance during our observing run. We are very grateful to J. Najita, J. Shields, J.R.Graham and W. van Breugel for useful comments on the manuscript. H.S. gratefully acknowledges NSF grant # AST-9225133. HS and MD thank KPNO for generous allotments of telescope time.

## REFERENCES

Baade, W. & Minkowski, R. 1954*a*, ApJ, 119, 206.
Baade, W. & Minkowski, R. 1954*b*, ApJ, 119, 215.
Baldwin, J.A., Phillips, M.M. & Terlevich, R. 1981, PASP, 93, 5.
Bennett, C.L., Lawrence, C.R., Burke, B.F., Hewitt, J.N. & Mahoney, J. 1986, ApJS, 61, 1.
Bruzual A., G. & Charlot, S. 1993, ApJ, 405, 538.
Burbidge, E.M. & Burbidge, G.R. 1959, ApJ, 129, 271.
Casali, M. & Hawarden, T. 1992, JCMT-UKIRT Newsletter No. 4, p.33.
Chambers, K.C., Miley, G.K., & van Breugel, W.J.M. 1987, Nature, 329, 604
Chambers, K.C., Miley, G.K., & van Breugel, W.J.M. 1990, ApJ, 363, 21
Charlot, S. & Fall, S.M. 1993, ApJ, 415, 580.
Charlot, S. & Fall, S.M. 1991, ApJ, 378, 471.
Chen, W.A. & Neufeld, D.A. 1994, ApJ, *in press.*
Colless, M., Schade, D., Broadhurst, T.J. & Ellis, R.S. 1994, MNRAS, 267, 1108
Colless, M., Ellis, R., Taylor, K. & Hook, R.N. 1990, MNRAS, 244, 408.
Djorgovski, S., Thompson, D. & Smith, J.D. 1993, in Texas/PASCOS '92, Relativistic Astrophysics and Particle Cosmology, C. Akerlof & M. Srednicki, ed., (Ann.N.Y.Acad.Sci.) 688, 515.
Eales, S.A. & Rawlings, S. 1993, ApJ, 411, 67.
Eales, S., Rawlings, S., Puxley, P., Rocca-Volmerange, B., & Kuntz, K. 1993, Nature, 363, 140.
Eisenhardt, P.R.M., & Dickinson, M. 1992, ApJ, 399, L47.
Elston, R., McCarthy, P.J., Eisenhardt, P., Dickinson, M., Spinrad, H., Januzzi, B.T. & Maloney, P. 1994, AJ, 107, 910.
Elston, R., Rieke, G.H., & Rieke, M. 1991, in *Astrophysics with Infrared Arrays,* ed. R. Elston (ASP, San Francisco), p.3.
Fall, S.M. & Pei, Y.C. 1993, ApJ, 402, 479.
Ferland, G.J. 1994, University of Kentucky Internal Report.
Ferland, G.J. & Netzer, H. 1983, ApJ, 264, 105.
Graham, J.R., Matthews, K., Soifer, B.T., Nelson, J.E., Harrison, W., Jernigan, J.G., Lin, S., Neugebauer, G., Smith, G., & Ziomkowski, C. 1994, ApJ, 420, L5.
Heckman, T.M., Smith, E.P., Baum, S.A., van Breugel, W.J.M., Miley, G.K., Illingworth, G.D., Bothun, G.D. & Balick, B. 1986, ApJ, 311, 526.
Hu, E.M., & Ridgway, S.E. 1994, AJ, 107, 1303.
Hummer, D.G. & Storey, P.J. 1987, MNRAS, 224, 801.
Hummer, D.G. & Kunasz, P.B. 1980, ApJ, 236, 609.
Kennicutt, R.C. 1992*a*, ApJ, 388, 310.
Kennicutt, R.C. 1992*b*, ApJS, 79, 255.
Kriss, G.A. *et al.* 1992*a*, ApJ, 392, 485.
Kriss, G.A., Davidsen, A.F., Blair, W.P., Ferguson, H.C. & Long, K.S. 1992*b*, ApJ, 394, L37.
Lacy, M. *et al.* 1994, MNRAS, *in press.*
Landolt, A.U., 1992, AJ, 104, 340.
Lilly, S.J. 1988, ApJ, 333, 161.
Lilly, S.J. 1993, ApJ, 411, 501.
Massey, P., Strobel, K., Barnes, J.V. & Anderson, E. 1988, ApJ, 328, 315.
McCarthy, P.J., 1993, ARA&A, 31, 639.
McCarthy, P.J., Elston, R. & Eisenhardt, P. 1992, ApJ, 387, L29.

McCarthy, P.J., Spinrad, H., van Breugel, W., Liebert, J., Dickinson, M., Djorgovski, S. & Eisenhardt, P. 1990, ApJ, 365, 487.
McCarthy, P.J., van Breugel, W.J.M., Spinrad, H. & Djorgovski, S. 1987, ApJ, 321, L29
Meier, D.L. *et al.* 1989, AJ, 98, 27.
Metcalfe, N., Shanks, T., Fong, R. & Jones, L.R. 1991, MNRAS, 249, 498.
Minkowski, R. 1960, ApJ, 132, 908.
Neufeld, D.A. 1990., ApJ, 350, 216.
Oke, J.B. 1990, AJ, 99, 1621.
Osterbrock, D.E. 1989, *Astrophysics of Gaseous Nebulae and Active Galactic Nuclei*, University Science Books.
Partridge, R.B. & Peebles, P.J.E. 1967, ApJ, 147, 868.
Rowan-Robinson, M. *et al.* 1991, Nature, 719, 351.
Savage, B.D. & Mathis, J.S. 1979, ARA&A, 17, 73.
Schneider, D.P., Schmidt, M. & Gunn, J.E. 1991, AJ, 101, 2004.
Schweizer, F. 1976, ApJS, 31, 313.
Solomon, P.M., Radford, S.J.E. & Downes, D. 1992, Nature, 356, 318
Spinrad, H., Dickinson, M., Schlegel, D. & Gonzalez, R. 1993, in *Observational Cosmology*, ed. G. Chicarini, A. Iovino, & D. Maccagni, *ASP Conf. Ser.* v.51, p.585.
Tresse, L., Hammer, F., LeFèvre, O., & Proust, D. 1993, A&A, 277, 53.
van Ojik, R., Röttgering, H.J.A., Miley, G.K., Bremer, M.N., Macchetto, F. & Chambers, K.C. 1994, A&A, in press.
Wade, C.M., Hjellming, R.M., Kellermann, K.I. & Wardle, J.F.C. 1971, ApJ, 170, L11.

## Figure Captions:

**Figure 1:** Keck LRIS $R$-band image of MG1019+0535. The cross marks our best estimate for the position of the radio core. The scale bar of $1''$ corresponds to 13.5 kpc at z=2.765 and 9.3 kpc at z=0.66 ($H_\circ$=50$\,$km$\,$s$^{-1}\,$Mpc$^{-1}$, $q_\circ$=0). The position of component A is roughly $\alpha_{1950} = 10^h16^m56\overset{s}{.}8$, $\delta_{1950} = +5°49'39''$. The $R$ magnitudes of A and B are 23.7 and 23.0 respectively.

**Plate 2:** Co-registered $R$ and $K_S$ images of the MG1019+0535 field. The extremely red object marked D in the $K_S$ image is undetected in our deep Keck $R$ image ($R-K>$7) where its location is denoted by the cross. The scale bar of $5''$ corresponds to 67.5 kpc at z=2.765 and 46.3 kpc at z=0.66 ($H_\circ$=50$\,$km$\,$s$^{-1}\,$Mpc$^{-1}$, $q_\circ$=0).

**Figure 3:** Contour map of the VLA 8.44GHz radio observation of MG1019+0535. The contours are drawn at (3,4,5,8,16,32,64)$\sigma$ where $\sigma = 0.22$mJy/beam. There are no negative contours below the $-3\sigma$ level within this $4''\times4''$ box. The extension seen to the south of the core component is not real.

**Figure 4:** Keck LRIS spectra of the two optical components of MG1019+0535. Note that whereas component A shows several strong emission lines at z=2.765, component B has only one moderately strong line that is close in observed wavelength to the HeII$\lambda$1640 line in A. The solid triangles mark the locations of H$\beta$ and [OIII]$\lambda$5007 under the assumption that the line at 6192Å is [OII]$\lambda$3727 at z=0.66; no line emission is detected at these locations. The turn down in the spectrum of B for $\lambda>$7300Å is not real and is due to instrumental problems.

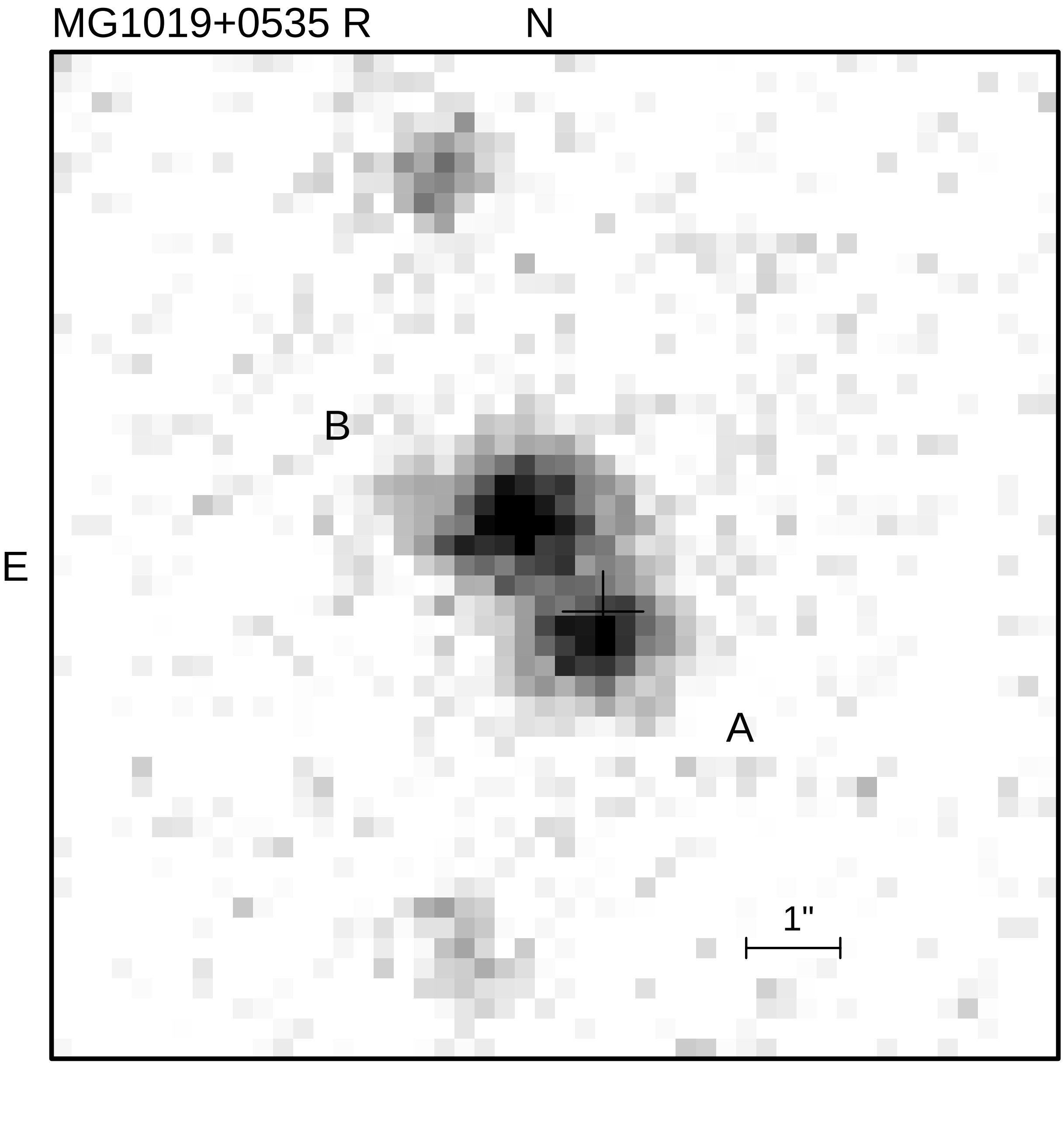

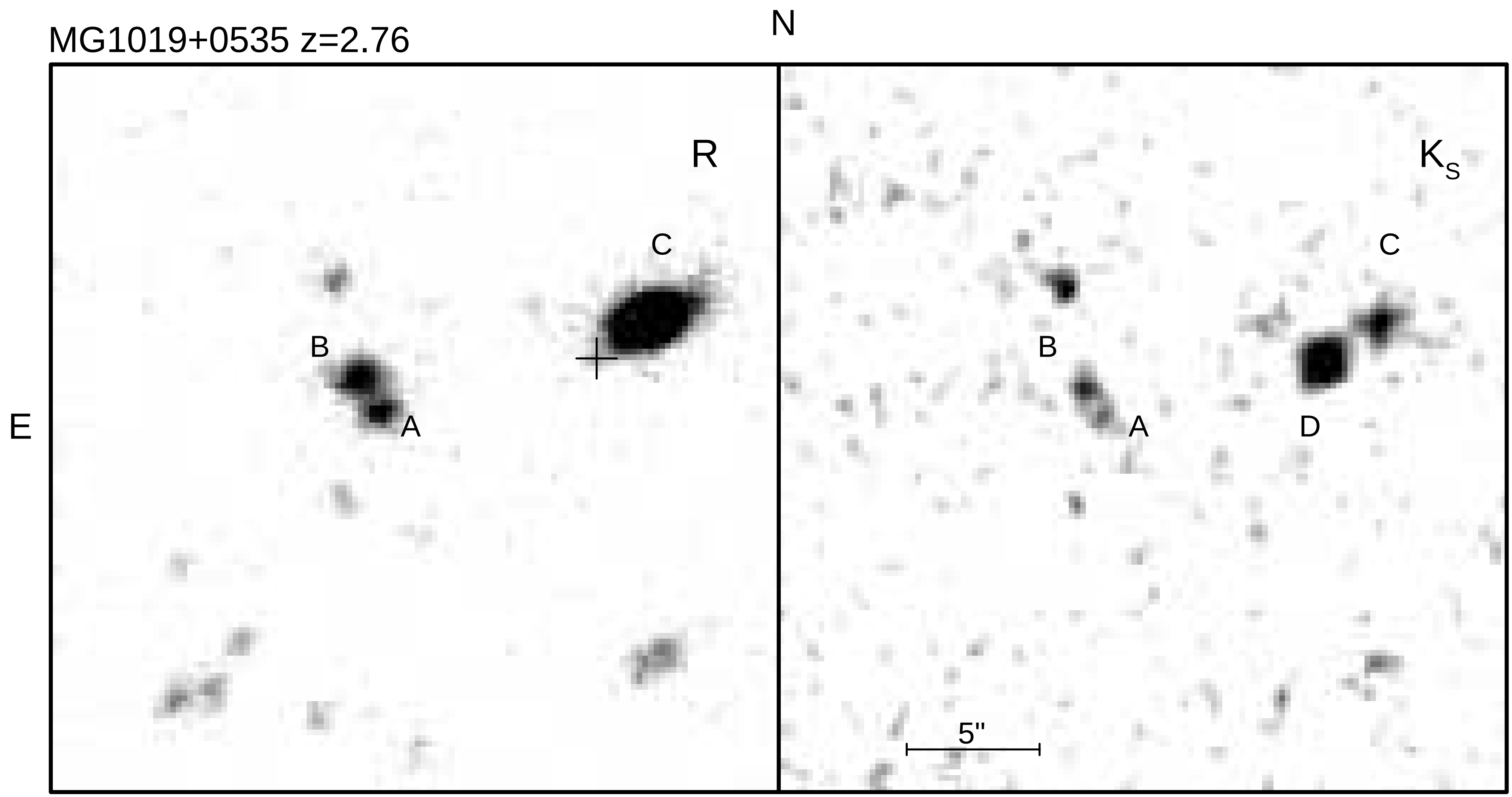

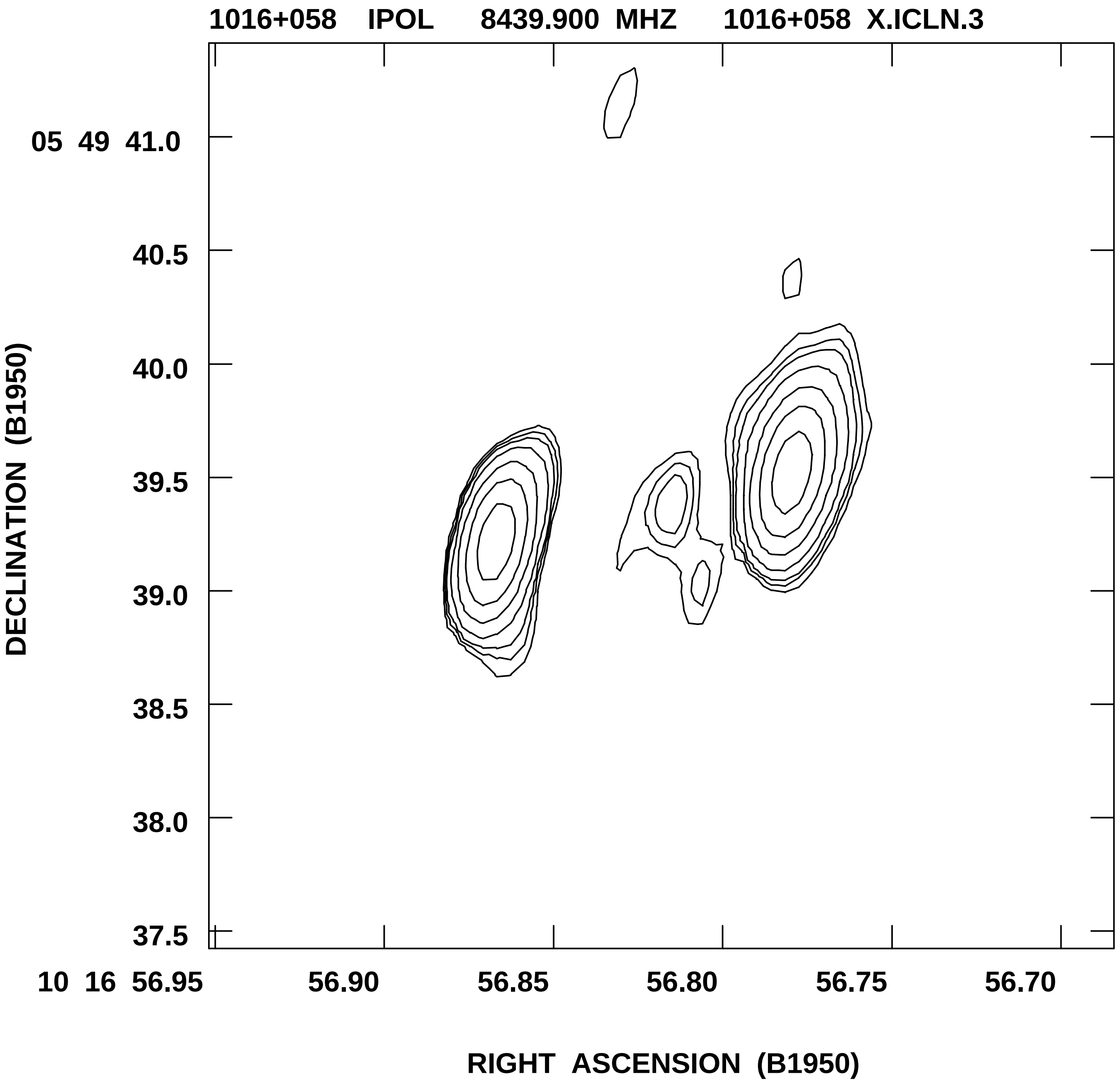

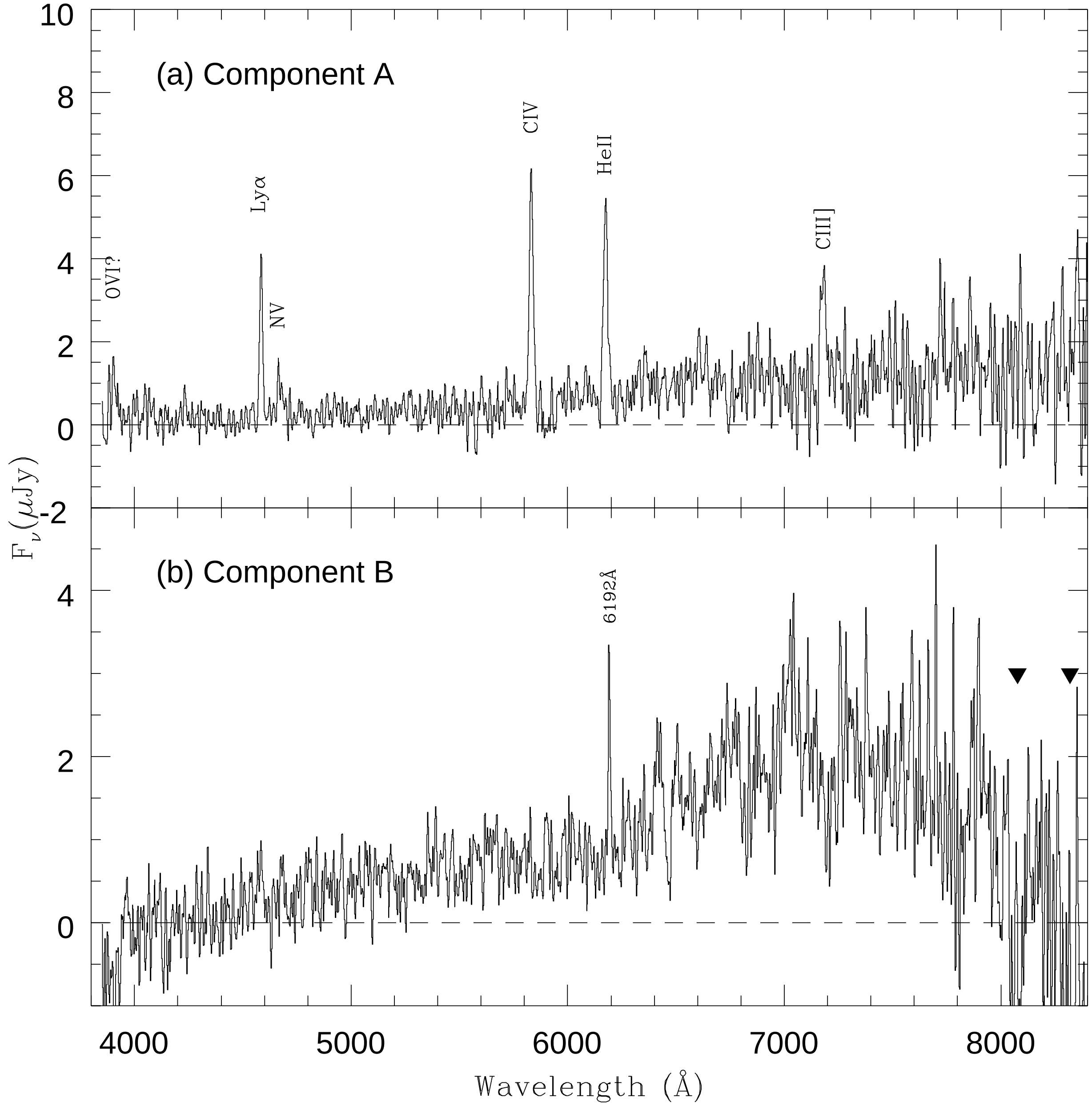

Table 1: Observed Radio and Optical Flux Densities

| Object | $\alpha_{1950}$, $\delta_{1950}$ | $\nu$ MHz | $F_\nu$ mJy | Comments |
|---|---|---|---|---|
| Radio Core | $10^h16^m56\overset{s}{.}815(\pm0\overset{s}{.}010)$, $+5°49'39\overset{''}{.}30(\pm0\overset{''}{.}15)$ | 8439 | 3.0±0.4 | VLA |
| East Lobe | $+0\overset{''}{.}78$, $-0\overset{''}{.}09$ | 8439 | 25.6±1.0 | VLA |
| West Lobe | $-0\overset{''}{.}54$, $+0\overset{''}{.}22$ | 8439 | 30.5±1.0 | VLA |
| Total Source | | 365 | 925±31 | Texas Catalog |
| | | 1400 | 454 | White & Becker 1992 |
| | | 1490 | 360±6 | VLA |
| | | 4850 | 132±19 | Gregory & Condon 1991 |
| | | 8439 | 59.1±1.5 | VLA |
| Component A | $10^h16^m56\overset{s}{.}82$ | $4.61\times10^8$ | $0.98\times10^{-3}$ | $R$=23.70±0.04 in 3″ dia. |
| | $+5°49'39\overset{''}{.}0$ | | | $R_c$=23.90 (corr. for emission lines) |
| | | $1.36\times10^8$ | $6.4\times10^{-3}$ | $K\approx20$ |
| Component B | $10^h16^m56\overset{s}{.}87$ | $4.61\times10^8$ | $1.79\times10^{-3}$ | $R$=23.05±0.02 in 3″ dia. |
| | $+5°49'40\overset{''}{.}2$ | $1.36\times10^8$ | $7.15\times10^{-3}$ | $K$=19.89±0.16 |
| A+B | | $4.61\times10^8$ | $3.73\times10^{-3}$ | $R$=22.25±0.02 in 7″ dia. |
| | | $1.36\times10^8$ | $14.9\times10^{-3}$ | $K$=19.09±0.24 |

Table 2: Observed Spectral Lines and Their Fluxes

| Object | $\lambda_{obs}$ Å | $F_{obs}$ $10^{-17}$erg/s/cm$^2$ | $W_\lambda^{obs}$ Å | FWHM$_{obs}$ Å | ID |
|---|---|---|---|---|---|
| Component A | 4584.3±0.3 | 8.4±0.6 | 268 | 13.5 | Ly$\alpha\lambda$1216 |
| | | | | | $L_{\mathrm{Ly}\alpha} = 1.6(\pm 0.1) \times 10^{43}$erg s$^{-1}$ |
| | 4664.8±0.8 | 2.3±0.6 | 66 | 10.6 | NV$\lambda$1240 |
| | 5831.9±0.4 | 10.4±0.7 | 257 | 19.6 | CIV$\lambda$1549 |
| | 6174.9±0.5 | 8.5±0.6 | 170 | 21.3 | HeII$\lambda$1640 |
| | 7179.6±1.9 | 4.9±1.0 | 93 | 27.7 | CIII]$\lambda$1909 |
| | 3903: | 3.8: | | 11: | OVI$\lambda$1034? |
| Component B | 6192.0±0.5 | 2.4±0.4 | 46 | 10.4 | [OII]$\lambda$3727? or HeII$\lambda$1640? |

Table 3: Comparison of MG1019+0535 Line Ratios with Other Objects

| Spectral Line | MG1019 +0535A | Average [a] Radio Galaxy | F10214 +4724[c] | TX0211 −122[d] | NGC 1068[e] | Model I[b] $\log U = -1.8$ | Model II[b] $\log U = -2.8$ |
|---|---|---|---|---|---|---|---|
| Ly$\alpha$ | 0.99[f] | 9.75 | $<$0.24 | 1.50 | 4.75 | 19 | 21 |
| NV | 0.27 | 0.48 | 1.86 | 1.33 | 1.32 | 0.024 | 0.0016 |
| CIV | 1.23 | 1.15 | 2.10 | 1.67 | 1.86 | 1.13 | 0.01 |
| HeII | 1.00 | 1.00 | 1.00 | 1.00 | 1.00 | 1.00 | 1.00 |
| CIII] | 0.58 | 0.56 | 0.57 | 0.50 | – | 0.67 | 0.16 |

[a] McCarthy (1993)

[b] Shields, pers. comm.

[c] Elston *et al.* (1994)

[d] van Ojik *et al.* 1994, 5″ aperture

[e] Kriss *et al.* (1992)

[f] The Ly$\alpha$/HeII$\lambda$1640 ratio is 0.69 if we correct for HeII$\lambda$1215 contamination using the Hummer and Storey (1987) Case B ratio of HeII$\lambda$1215/HeII$\lambda$1640 for $N_e$=$10^4$cm$^{-3}$ and $T$=$10^4$K.